\documentstyle[11pt,aaspp4]{article}
\newcommand{\etal}{{\it et al.\ }}

\begin{document}

\title{\ion{Ca}{2} Triplet Spectroscopy of Giants in SMC Star Clusters: Abundances,
Velocities and the Age-Metallicity Relation}

\author{G. S. Da Costa}
\affil{Mount Stromlo and Siding Spring Observatories, The Australian National
University, Private Bag, Weston Creek Post Office, ACT 2611, Australia\\
Electronic mail: gdc@mso.anu.edu.au}
\authoremail{gdc@mso.anu.edu.au}

\and

\author{D. Hatzidimitriou}
\affil{Physics Department, University of Crete, P.O. Box 2208, 
Heraklion 710 03, Crete, Greece\\
Electronic mail: dh@physics.uch.gr}
\authoremail{dh@physics.uch.gr}

\begin{abstract}

We have obtained spectra at the \ion{Ca}{2} triplet of individual red giants in
seven SMC star clusters whose ages range from $\sim$4 to 12 Gyr.  The spectra
have been used to determine mean abundances for six of the star clusters 
to a typical precision of 0.12 dex.  When combined with existing data
for other objects, the resulting SMC age-metallicity
relation is generally consistent with that for a simple model of chemical
evolution, scaled to the present-day SMC mean abundance and gas mass
fraction.  Two of the clusters (Lindsay~113 and NGC~339), however, 
have abundances that are $\sim$0.5 dex lower than that expected from the
mean age-metallicity relation.   It is suggested that the formation of 
these clusters, which have ages of $\sim$5 Gyr, may have involved 
the infall of unenriched gas, perhaps
from the Magellanic Stream.  The spectra also yield radial velocities 
for the seven clusters.  The resulting velocity dispersion is 16~$\pm$~4 
km~s$^{-1}$, consistent with those of the SMC planetary nebula and carbon 
star populations. 

\end{abstract}


\vspace{6mm}
\begin{center}
{\large Accepted for Publication in {\it The Astronomical Journal}}
\end{center}

\pagebreak

\section{Introduction}

The age-metallicity relation of a stellar system is one of 
the fundamental diagnostics through
which we can learn about the chemical enrichment processes that occur
during the system's evolution.  Within the solar neighbourhood
for example, it is possible to determine the age-metallicity relation for
the Galactic disk from the ages and metallicities of field stars 
(e.g., \markcite{EA93}Edvardsson \etal 1993), and modelling
of this relation and its scatter suggests a complex evolution involving gas
inflow and the radial diffusion of orbits in the presence of a radial
abundance gradient (e.g., \markcite{PT95}Pagel \& Tautvaisiene 1995).  However,
for more distant systems, we can no longer use individual F- and G-type dwarfs
to determine ages and abundances.  Instead we must use star clusters since,
at least in principle, it is relatively straightforward to determine a
cluster's age and abundance.  In particular, for our nearest galactic
neighbors it is possible to constrain cluster ages via main sequence turnoff
luminosities even if the clusters are as old as the galactic halo globular
clusters.  Thus provided clusters of all ages can be found, which is 
of course a function of formation and destruction processes, the age-metallicity
relations of these nearby systems can be fully outlined.  

It is with this proviso, however, that a major difficulty often arises.  For
example, in the Large Magellanic Cloud (LMC) there is a well established
population of old metal-poor star clusters whose properties are comparable
to the galactic halo globular clusters 
(e.g., \markcite{SS92}Suntzeff \etal 1992).  There is also a large population
of intermediate-age clusters, but the ages of these clusters are all less
than 3 -- 4 Gyr (e.g., Da~Costa 1991, Olszewski \etal 1996, 
\markcite{DG97}Geisler \etal 1997) and only a single cluster, ESO~121-SC03
(\markcite{MH86}Mateo \etal 1986) is known to fall in the ``age gap''
between the old and intermediate-age objects.  As emphasized by
Olszweski \etal (1996), for example, this gap in the LMC cluster 
age distribution also
represents an ``abundance gap'', in that the old clusters are all metal-poor
while the intermediate-age clusters are all relatively metal-rich
(Olszewski \etal 1991), approaching even the present-day abundance in the LMC\@.
Yet because of the lack of clusters older than $\sim$3 Gyr, we have essentially
no constraints on the timescale for what was the major enrichment event 
in the LMC's history.
The Small Magellanic Cloud (SMC), on the other hand, is known to have a
different distribution of cluster ages from the LMC (e.g., Da~Costa 1991).
In particular, while the number of clusters with ages determined from main
sequence turnoff photometry is smaller than for the LMC, there is no sign
in these data for any substantial ``age gap''.  Thus, in principle, the SMC
age-metallicity relation defined by the star clusters can cover from the 
present day back almost to the age of the oldest galactic halo globular
clusters\footnote{NGC~121, the oldest SMC star cluster for which age 
information exists, is perhaps 2~--~3~Gyr younger than the galactic halo
globular clusters (e.g., Stryker \etal 1985).}.  

The SMC age-metallicity relation, as it stood prior to the present work
(e.g., Stryker \etal 1985, Da~Costa 1991, Olszewski \etal 1996) 
had fundamentally two components.
First, it appeared that following a presumed initial burst of star formation
that brought the cluster abundances up to [Fe/H] $\approx$ --1.3, the
subsequent net rate of enrichment was very low until perhaps 2 -- 3 Gyr ago.
Second, commencing at an age of $\sim$2 -- 3 Gyr, the rate of enrichment
apparently increased to bring the cluster abundances up to the present-day SMC
abundance of [Fe/H] $\approx$ --0.6 dex.  This sort of age-metallicity 
relation is very different from that of the solar neighbourhood and it also
differs substantially from that expected from the simple model of chemical
evolution.  As summarized by Dopita (1991), the first component of this 
relation could be explained if the infall rate of unenriched gas matched
the rate of chemical enrichment, or if the newly synthesized metals were
lost in a strong galactic wind (or a combination of these processes).  The
second component could result from the cessation of these processes, or from
an increased star formation rate perhaps as a result of an
interaction with the LMC\@.  However, it must be emphasized that the
majority of the cluster abundances used to define this age-metallicity
relation are rather uncertain: most are derived from the color of the 
cluster red giant branch in the color-magnitude diagram.  This technique
depends on the reddening of the cluster and on the zeropoint of the photometry,
and it also 
contains the uncertainty inherent in the calibration of a parameter that is
sensitive to both age and abundance.  

It is obvious then that to verify the chemical history of the SMC implied by 
this age-metallicity relation, it needs to be redetermined
using a consistent, accurate and reddening independent method.  Spectroscopy
at the \ion{Ca}{2} triplet of globular cluster red giants analyzed in
terms of the magnitude difference from the horizontal branch is such a
method (see, for example, Da~Costa \& Armandroff 1995, Rutledge \etal 1997
and the references therein), and as Olszewski \etal (1991) have shown, it
can be extended to red giants in intermediate-age star clusters since the
higher gravity of the younger stars has little effect, at least for ages
in excess of perhaps 2 Gyr.  
The purpose of this paper then is to present \ion{Ca}{2} triplet spectroscopy
for red giants in a number of SMC star clusters and to discuss the 
implications of these new results for the SMC age-metallicity relation and
the SMC chemical history.  Spectroscopy of
individual stars has the added bonus that we can also determine the cluster
radial velocities.  The kinematics of the LMC cluster system has produced
some intriguing results (e.g., Schommer \etal 1992) and while the sample
of clusters observed here is small, the velocity results can nevertheless 
be compared
with the kinematics of other SMC populations.

The outline of the remainder of the paper is as follows: in the following 
section the sample of SMC clusters observed, the data reduction procedure,
and the line strength and radial velocity analysis techniques are described.  
Section \ref{abund} discusses in some detail the derivation of abundances
from the observed line strengths.  Then, in the first part 
of \mbox{Sect.\ \ref{results}},
the kinematics implied by the cluster radial velocities are derived and
compared with those of other SMC populations with similar ages.  In 
the second part
of this section (Sect.\ \ref{amr}) the full SMC age-metallicity relation 
is constructed.  The
implications of this relation for the chemical evolution of the SMC are
discussed in the final section.

\section{Observations and Reductions}

\subsection{The SMC Cluster Sample} \label{samp}

Compared to the LMC, there are fewer well-studied star clusters in the SMC,
though as noted above, the SMC clusters do not show the large ``age-gap''
exhibited by the LMC cluster population.  For this reason we initially
selected only those clusters with both good quality color-magnitude (c-m) 
diagrams and ages, as determined from main sequence turnoff 
photometry, greater than
$\sim$2 Gyr.  There are six such clusters: 
Lindsay~1 (\markcite{OS87}Olszweski \etal 1987),
Kron~3 (\markcite{RD84}Rich \etal 1984), 
Lindsay~11 (\markcite{MJ92}Mould \etal 1992), 
NGC~121 (\markcite{SD85}Stryker \etal 1985), 
NGC~339 (Olszewski, 1994, priv.\ comm.), and 
Lindsay~113 (\markcite{MD84}Mould \etal 1984).  
For each cluster the brighter half dozen or so candidate cluster red giants
were selected for spectroscopic observation at the \ion{Ca}{2} triplet.  In
this selection process we were careful to ensure that
none of the stars chosen had been previously identified via IR-photometry
and/or low resolution spectroscopy as upper-AGB stars (whether of C, M or S
spectral type).  Then, in order to increase the sample, we 
constructed a second list of SMC clusters for which main sequence turnoff
color-magnitude diagrams were not available, but for which independent 
data suggested an
age in the $\gtrsim$2 Gyr to 12 Gyr range.  The only cluster in this second
list for which stars were actually observed was NGC~361.  
\markcite{HA58}Arp (1958) gives a 
photographic c-m diagram for this cluster but it does not reach as faint 
as even the expected magnitude for the core helium burning stars.  It 
therefore yields little age information.  However, when corrected for the
effects of a superposed bright star 
(cf.\ \markcite{VB81}van den Bergh 1981) the integrated $UBV$
colors of NGC~361 are similar to those of the six clusters in the
first list.  Thus it is likely that NGC~361 is of intermediate-age, 
and the brighter red giants from the c-m diagram of
Arp (1958) were added to the sample for spectroscopic study.

\subsection{Observations}

\begin{figure}
\epsscale{0.8}
\plotone{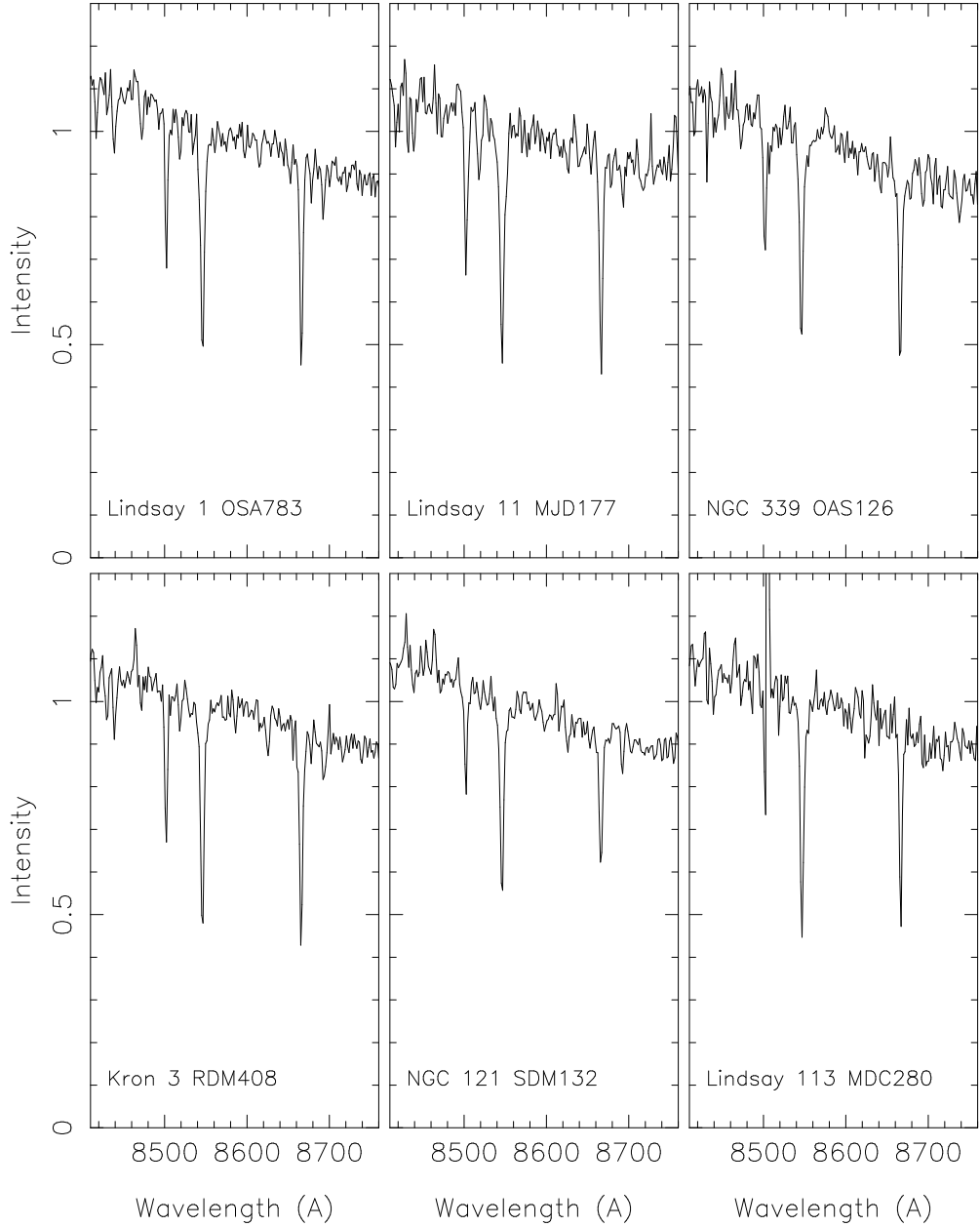}
\caption{Examples of observed spectra 
for red giants in six SMC star
clusters.  The most prominent spectral features are the lines of the \ion{Ca}{2}
triplet at $\lambda$8498\AA, $\lambda$8542\AA\ and $\lambda$8662\AA\@.  Each
spectrum has been normalized to unity at $\sim$8600\AA\@.  The 
stars are identified in the panels.  Note that the apparent emission feature
near $\lambda$8498\AA\ in the spectrum of Lindsay~113 MDC280 is a cosmic ray
artifact. \label{spectra}}
\end{figure}

The program stars were observed with the Anglo-Australian Telescope using
the RGO spectrograph with the 25cm camera and a Tek 1k CCD, during 
runs in 1992 September (second halves of five nights) and 1994 August (second
halves of two nights).  The instrumental setup was identical to that 
discussed in \markcite{\DA95}Da~Costa \& Armandroff 
(1995, hereafter DA95) which can be 
consulted for a more detailed description.  In brief, the spectra cover the
wavelength interval 
$\sim$7700--9300\AA\  at a resolution of $\sim$3\AA\@.  Total exposure times 
for the
SMC cluster stars ranged from 800 seconds to 2 hours for the faintest stars.  
Wherever possible two stars were observed simultaneously by rotating the
spectrograph slit to an appropriate position angle.  Standard 
IRAF\footnote{IRAF is distributed by the National Optical Astronomy
Observatories, which is operated by the Association of Universities for
Research in Astronomy, Inc.\ (AURA) under cooperative agreement with the
National Science Foundation.} procedures were used to generate sky-subtracted
wavelength-calibrated spectra from the raw data.  Some examples 
of the final spectra are
shown in Fig.\ \ref{spectra}.  Spectra of galactic globular cluster red giants
obtained during these observing runs are discussed in DA95, who have
demonstrated that the data from the two observing runs have no systematic
differences and can be used interchangably.  At this point it is also worth
mentioning that Lindsay 11 star 1 from 
\markcite{MA82}Mould \& Aaronson (1982) was also
observed during the 1994 August run. This star has the infra-red colors of
an upper-AGB carbon star (cf.\ Mould \& Aaronson 1982) and the presence 
of strong CN-bands at $\lambda \approx 7850$\AA\ and 8050\AA\ in the
spectrum confirm this classification.  Such bands were not seen in the spectrum
of any other star observed.  There were also no indications in these spectra of
TiO bands whose presence can affect the measurement of the \ion{Ca}{2}
triplet line strength (e.g., Olszewski \etal 1991).

\subsection{Line Strength Measurements}

The (pseudo) equivalent widths of the two stronger lines of the \ion{Ca}{2}
triplet, $\lambda$8542\AA\ and $\lambda$8662\AA, were measured from the
red giant spectra by
applying the gaussian fitting technique of DA95 (see also 
\markcite{AD91}Armandroff \& Da~Costa 1991).  The sum of these line strength
measures, denoted by W$_{8542}$+W$_{8862}$, was then computed for each 
observation.  Final values for 33 stars in the fields of the 7 SMC
clusters are given in Table \ref{data_tab}.  In this table
the errors accompanying the W$_{8542}$+W$_{8862}$ values are those which result
from the uncertainties in the parameters that define the gaussian line profile 
fits.  Comparison of these errors with the mean single observation error in
W$_{8542}$+W$_{8862}$ derived from the 31 repeat observations of 15 stars shows
that these gaussian line profile fit uncertainties probably overestimate 
the true errors by perhaps 25 percent (0.32\AA\ versus 0.40\AA).  Nevertheless,
we have been conservative and retained the individual gaussian line profile
fit uncertainties.

For four of the seven clusters, the available photometry is in the $B$ and $R$
bandpasses, whereas the abundance calibration procedure (cf.\ DA95)
requires $V$ magnitudes for both the individual stars and for the 
horizontal branch luminosity.  To overcome this situation we have made
use of the empirical transformation:
\begin{equation}
B-V = (0.685 \pm 0.005)(B-R) - (0.06 \pm 0.01)
\end{equation}
This transformation is based on high quality photoelectric $BVR$ photometry 
of red giants in the galactic globular clusters 47~Tuc, NGC~288, NGC~362,
NGC~6397 and NGC~6752 (E.~M.~Green, 1987, priv.\ comm.) and is appropriate
for the reddening E(B--V) = 0.04 mag assumed for the SMC clusters.  
The rms deviation
about the fitted line is only 0.010 mag and the equation is valid between
$B-R \approx 1.1$ $(B-V \approx 0.7)$ and 
$B-R \approx 2.5$ $(B-V \approx 1.65)$.
The $V$ magnitude for each program star then follows from the observed $B$
magnitude and the computed $B-V$ color.  The same process has also been used to
fix the $V_{HB}$ values for these clusters: the median $B-R$ color of the
red clump of core helium burning stars (which constitutes the horizontal
branch in these clusters) was converted to $B-V$ via equation (1) and
combined with $B_{HB}$, taken as $R_{HB}$ plus the median $B-R$ color, to
generate $V_{HB}$.  This process can be verified for two of the four
clusters involved since independent $BV$ photometry is available.  For
Kron~3, the c-m diagram of 
\markcite{BG80}Gascoigne (1980) suggests $V_{HB}$ $\approx$ 19.3 while
\markcite{AL96}Alcaino \etal (1996) list $V_{HB}$ = 19.50 $\pm$ 0.05.  Both
these values are
in agreement with our adopted value of 19.44 derived from the
$(R, B-R)$ photometry of Rich \etal (1984).  We note also that the Lindsay~1 
c-m diagram in Gascoigne (1980) gives $V_{HB}$ $\approx$ 19.3, again
in good accord with our value $V_{HB}$ = 19.20 adopted from the ($V$, $B-V$)
photometry of Olszewski \etal (1987).  For NGC~121, the
photometry of \markcite{WT63}Tifft (1963) for zones 2 and 3,
which should be free of
photographic background effects (cf.\ Stryker \etal 1985), gives
$V_{HB}$ $\approx$ 19.6.  This agrees well with the value $V_{HB}$ = 19.60
derived from the $(R, B-R)$ photometry of Stryker \etal (1985).  The adopted
$V_{HB}$ values for all the sample clusters (except NGC~361 for which no
suitable photometry exists) are given in Table \ref{results_tab}.

\begin{figure}
\epsscale{0.65}
\plotone{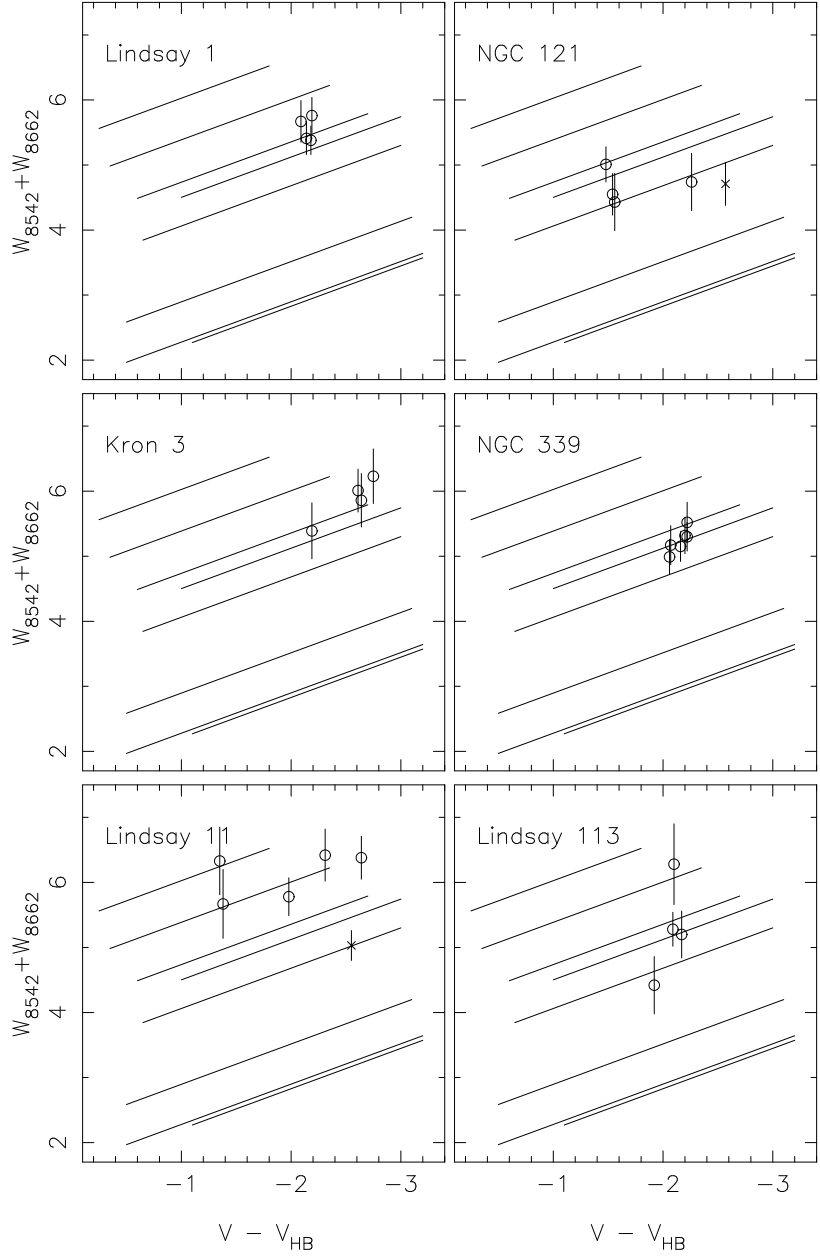}
\caption{Plots of the \ion{Ca}{2} line strength index
W$_{8542}$+W$_{8662}$ in \AA\ against magnitude difference from the
horizontal branch $V-V_{HB}$ for individual stars in the fields of six SMC
star clusters.  Probable cluster members are plotted as open circles while
probable non-members are shown as $\times$-symbols.  The lines in the panels 
are taken from Da~Costa \& Armandroff (1995) and represent
the \ion{Ca}{2} line strength, magnitude relations for galactic globular
clusters.  In order of increasing \ion{Ca}{2} line strength, the clusters
are M15, NGC~4590 (M68), NGC~6397, NGC~6752, M5, M4, 47~Tuc and NGC~5927.
These clusters have abundances [Fe/H] on the Zinn \& West (1984) scale of
--2.17, --2.09, --1.91, --1.54, --1.40, --1.28, --0.71 and --0.31 dex, 
respectively. \label{w_vs_v}}
\end{figure}

In Fig.\ \ref{w_vs_v} the line strength measures are plotted against the 
$V$ magnitude difference from the horizontal branch, $V-V_{HB}$, for all
stars observed in the fields of 6
of the 7 clusters in the sample.  The lack of a c-m diagram
reaching faint enough to reveal the magnitude of the horizontal branch (or
red clump) in NGC~361 precludes analysis of the line strengths for this
cluster.  The galactic globular cluster calibration lines in this Figure
have been taken from DA95 and the panels of this diagram provide 
one indication of
the cluster membership status for each of the stars observed.  Since there
is no reason to suspect that any of these clusters contain a substantial
internal abundance range, we can assume that any star whose \ion{Ca}{2}
triplet line strength lies significantly away from those of
the majority, is not likely to be a member of that
cluster.  Inspection of Fig.\ \ref{w_vs_v} shows that the
star MJD240 in the field of Lindsay~11 has such a discrepant line strength,
while all other stars appear, on this basis at least, 
to be members of their respective clusters.  An additional, and independent,
means to investigate cluster membership status is provided by the
comparatively large heliocentric velocity of the SMC\@.  The measurement of
radial velocities from the observed spectra will be discussed in 
Sect.\ \ref{rv},
but for the moment we note that all stars except Lindsay 11 MJD240 and
NGC~121 SDM029 have radial velocities consistent with not only SMC membership
but also membership in their respective clusters.  Lindsay 11 MJD240 is 
apparently an SMC field
star while the low radial velocity of NGC~121 SDM029 indicates that
this star is probably a foreground Galactic star.  This star, and Lindsay~11
MJD240, will not be considered further.

As shown by a number of authors (e.g., DA95, 
\markcite{RH97}Rutledge \etal 1997, and the references therein), the slope
of the relation between the \ion{Ca}{2} line strength index
W$_{8542}$+W$_{8862}$ and $V-V_{HB}$ is independent of abundance, at least
for $V-V_{HB}$ $\lesssim$ --0.5 mag.  Thus the individual line strength
measures for the stars observed in a cluster can be combined into a single
quantity, the reduced equivalent width, denoted by W$^{\prime}$.  W$^{\prime}$
can then be calibrated in terms of metal abundance.   On the system of DA95
which we follow here, W$^{\prime}$ is defined as the mean value of the
quantity W$_{8542}$+W$_{8862}$+0.62($V-V_{HB}$) for the stars observed in a
particular cluster.  Since we have now established the cluster membership 
status of the program stars, we can compute W$^{\prime}$ for each cluster.
The resulting values are listed in Table \ref{results_tab}.  We 
note that for the
clusters Lindsay~1, Kron~3 and NGC~339, the cluster W$^{\prime}$ value 
is simply the mean of the stellar values 
since the individual W$_{8542}$+W$_{8862}$ errors are
all similar.  For the other three clusters, however, this is not the case and
the value of W$^{\prime}$ was computed as a weighted average, with the
weights taken as the inverse square of the W$_{8542}$+W$_{8862}$ errors
given by the gaussian line profile fit uncertainties for each individual star.
We note also that the error listed with the W$^{\prime}$ values in 
Table \ref{results_tab} contains not only the error, as expressed by the
standard error in the mean, from the scatter of
the individual values, but also includes the effect of an assumed $\pm$0.10 mag
uncertainty in the value of $V_{HB}$ for each cluster.  

For the three clusters
where the value of W$^{\prime}$ was computed as a weighted average, it is of
interest to see if the values change significantly if different weighting
schemes are adopted.  Consequently, we have computed values of W$^{\prime}$
for these clusters
for the case of no weighting, and for a weighting scheme in which the observed
mean single observation error in W$_{8542}$+W$_{8862}$ (0.32 \AA) 
together with the
number of observations per star was used to compute the weights.  We find that
in each of these cases, the values of W$^{\prime}$ computed 
for the three clusters
agree with those given in Table \ref{results_tab} to well within the
uncertainties listed in the Table.  The uncertainties in the alternate
W$^{\prime}$ values are also comparable to those given in 
Table \ref{results_tab} except for the case of Lindsay 113 and no weighting.
Here the outliers (cf.\~Fig.\ \ref{w_vs_v}), whose larger errors are ignored
in this particular calculation, generate an uncertainty that is 
a factor of two larger
than that listed in Table \ref{results_tab}.  We believe, however, that
the values listed in Table \ref{results_tab} are the appropriate ones to
adopt.  The relation between
these W$^{\prime}$ values and abundance will be discussed in 
Section \ref{abund}.

\subsection{Radial Velocities} \label{rv}

The determination of radial velocities from the spectra obtained in these
observing runs has been discussed in detail in DA95.  Briefly, the program
star spectra were cross-correlated with high S/N spectra of bright G- and 
K-type giant radial velocity standards.  The velocity zeropoint for each run
was set by minimizing, after applying heliocentric corrections, the 
difference between the observed and standard velocities for both the radial
velocity standards and a number of galactic globular cluster stars with well
determined velocities (see DA95 for details).  The resulting uncertainty in
the zeropoint of the velocity system is $\sim$2~km~s$^{-1}$.  Based on the
repeat observations, the individual radial velocity determinations for our
program stars have an average error of 6~km~s$^{-1}$.  

In Table \ref{results_tab} we give mean heliocentric velocities 
for the 7 SMC star
clusters observed.  These means were calculated from the individual cluster
star observations and the tabulated errors are the standard deviations of the
mean for the samples.  An external check on these velocities is provided by
the cluster NGC~121 for which other radial velocity determinations exist.
\markcite{ZW84}Zinn \& West (1984) find 
139 $\pm$ 20 km~s$^{-1}$ (after a zeropoint shift of --5 km~s$^{-1}$, 
see Hesser \etal 1986) while \markcite{HS86}Hesser \etal (1986) 
list V$_{\rm r}$ = 138 $\pm$ 15 km~s$^{-1}$.  Both determinations 
are based on integrated spectra of the cluster.
\markcite{SK86}Suntzeff \etal (1986) give V$_{\rm r}$ = 126 $\pm$ 15 km~s$^{-1}$
for a single red giant believed to be a member of NGC~121, while more recently,
\markcite{DM97}Dubath \etal (1997) have determined 
V$_{\rm r}$ = 147 $\pm$ 2 km~s$^{-1}$ for
NGC~121, again from an integrated cluster spectrum.  All these values are in
good accord with our determination for this cluster of 
V$_{\rm r}$ = 138 $\pm$ 4 km~s$^{-1}$, based on the velocities of 
four cluster red giants.  The kinematics of this sample of old and 
intermediate-age clusters will be discussed in Sect.\ \ref{motions}.

\section{Cluster Abundances} \label{abund}

In their paper DA95 give a calibration of W$^{\prime}$ with abundance using
[Fe/H] values for galactic globular clusters that have their origin
with Zinn \& West (1984).  There is an excellent correlation between
W$^{\prime}$ and [Fe/H]$_{\rm ZW84}$ though it is not linear over the 
entire [Fe/H]$_{\rm ZW84}$ range exhibited by the  
clusters.  DA95 chose to model the relation by two linear segments
that join at W$^{\prime}$ = 3.8\AA\ or [Fe/H]$_{\rm ZW84}$ = --1.44, finding
a rms dispersion about the fitted lines of $\leq$0.09 dex.  There are, however,
a number of issues that must be addressed before we can convert the 
SMC cluster W$^{\prime}$ values in Table~\ref{results_tab} to abundance,
using a calibration such as that of DA95.

The first of these is relatively straightforward.  The definition of
W$^{\prime}$ uses the difference between the apparent magnitude of the
red giant and that of the horizontal branch.  In the DA95 analysis, for
example, the $V_{HB}$ values used are those of globular cluster core helium
burning stars, i.e., stars whose age is $\sim$15 Gyr.  The majority of the
clusters studied here, however, have ages considerably younger than the
galactic globular clusters.  Now, as shown by, for example, the isochrones
of \markcite{BC94}Bertelli \etal (1994), at fixed abundance the luminosity
of core helium burning stars becomes progressively brighter as age
decreases or, more physically, as turnoff mass increases.  For example,
at an abundance of Z=0.001 (log Z/Z$_{\sun}$ = --1.3), the initial M$_{V}$
of core helium burning stars at an age of 4 Gyr is 0.28 mag brighter
than it is at 15 Gyr (Bertelli \etal 1994).  Thus the observed $V-V_{HB}$
value for a red giant in a younger cluster is smaller than it would be in a
cluster as old as the galactic globulars.  Failure to allow for this effect
would give rise to an overestimate of the true cluster abundance with the
difference systematically increasing with decreasing cluster age.  
Consequently, for a given abundance calibration (see below), we will tabulate
two abundances for each cluster, one based on the observed value of 
W$^{\prime}$ and a second derived from a corrected value of W$^{\prime}$.
The correction to W$^{\prime}$ comes from using the adopted cluster ages 
given in Table \ref{results_tab} with Bertelli \etal (1994) isochrones
of appropriate abundance
to estimate the change in $V_{HB}$.  An independent calculation
of this effect using the models of \markcite{SD87}Seidel \etal (1987b) in
conjunction with the Revised Yale Isochrones (\markcite{GD87}Green \etal 1987),
as described in \markcite{DH91}Hatzidimitriou (1991), 
gives essentially identical results.

The second issue involves the fact that W$^{\prime}$ is a {\it calcium} line
strength index whereas the calibration is to [Fe/H] values.  As long as
there is no difference in [Ca/Fe] between the program and calibration
objects, this is not a matter for concern.  However, calcium, being an
$\alpha$-element, has its origin primarily in SNeII and is enhanced, relative
to the solar ratio, in both globular cluster and metal-poor galactic halo stars
(e.g., \markcite{WS89}Wheeler \etal 1989).  Iron enrichment, on the
other hand, is thought to occur primarily from SNeIa which have a longer
evolutionary time than SNeII\@.  Thus the [Ca/Fe] ratio is quite likely to
be a function of star formation and enrichment history.  In particular, a
star in an SMC cluster with an age half that, or less, of the galactic 
globular clusters is not likely to have the same [Ca/Fe] value as a 
galactic globular cluster star, even though they may have the same [Fe/H].
At the present time we can do little about this issue other than to note 
first, that
investigation of element abundance ratios in SMC and LMC intermediate-age and 
old cluster and field stars is likely to be a prime science project for the
coming 8m class telescopes in the southern hemisphere, and second, that 
in young SMC field stars the [$\alpha$/Fe] ratio is approximately solar
despite the moderate overall Fe depletion (e.g., Hill 1997, Luck \& Lambert 
1992).  As regards our own data, two points can be made: first,
interpreting W$^{\prime}$ with a galactic globular cluster based relation
(i.e., one for which [Ca/Fe] $\approx$ +0.3) will lead to an
{\it underestimate} of the true [Fe/H] if [Ca/Fe] $\leq$ 0.3
in the program stars.  In such a case the derived [Fe/H] values will represent
lower limits on the actual abundance value.  Second, it would seem unlikely
that clusters with comparable ages could have substantially different
[Ca/Fe] ratios.  Thus the difference in W$^{\prime}$ between Lindsay~11 and
NGC~339, for example, is likely to reflect a significant difference in
overall abundance.

The final issue is that of the calibration relation itself.  As noted above,
in DA95 the W$^{\prime}$ calibration is tied to the globular cluster
abundance scale of Zinn \& West (1984).  In a recent paper, however, 
\markcite{CG97}Carretta \& Gratton (1997) have compiled a 
consistent set of [Fe/H] values derived from high dispersion spectroscopy
for galactic globular clusters.  These authors find that their [Fe/H] values
do not scale linearly with those of Zinn \& West (1984).  While
there is reasonable agreement for both the metal-poor ([Fe/H] $\approx$ --2.0)
and the metal-rich ([Fe/H] $\approx$ --0.7) clusters, at intermediate abundances
there are substantial differences (Carretta \& Gratton 1997).  For example,
the ``standard clusters'' M3, M13 and M5 have abundances on the Zinn \& West
(1984) scale of --1.66, --1.65 and --1.40 dex but have 
[Fe/H]$_{\rm CG97}$ = --1.34, --1.39 and --1.11, respectively, a difference of
a factor of 2 in Z\@.  

\begin{figure}
\epsscale{0.75}
\plotone{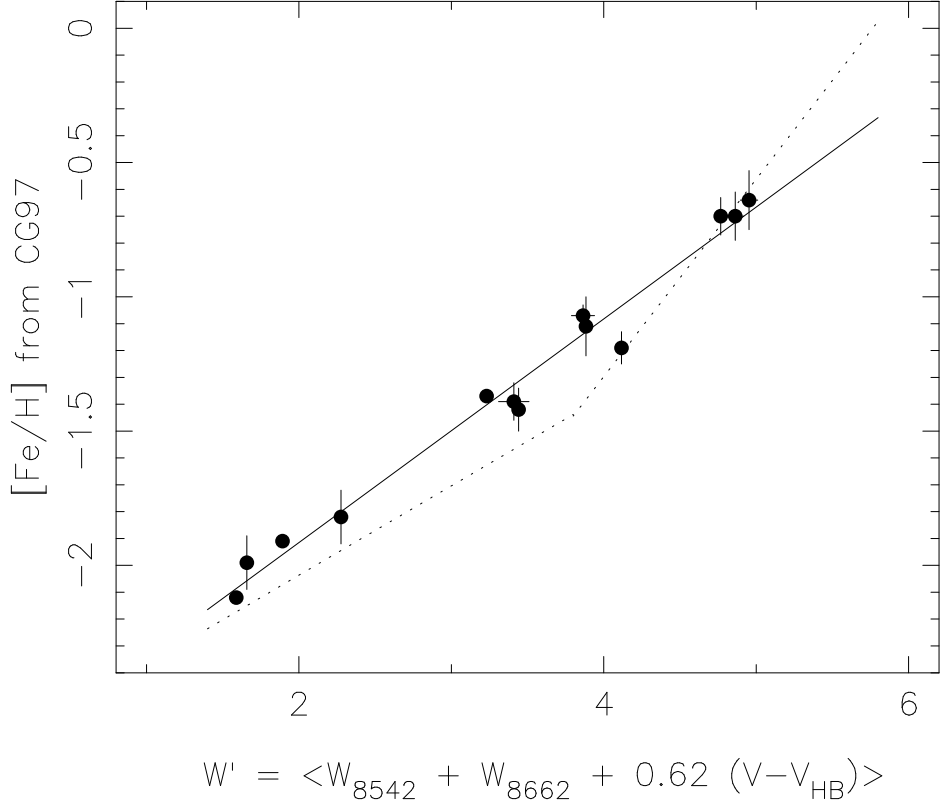}
\caption{The reduced equivalent 
width W$^{\prime}$ in \AA\ is plotted
against iron abundance [Fe/H] from Carretta \& Gratton (1997),
if available,
for the abundance calibration clusters of Da~Costa \& Armandroff (1995).
These [Fe/H] values all have their basis in high dispersion spectroscopy.
The solid line is a least squares fit to the data.  The
dotted line is the two linear segment fit of DA95 to [Fe/H] values on the
scale of Zinn \& West (1984).  Note that for W$^{\prime}$ $\approx$ 3.8\AA, 
the relations give [Fe/H] values that differ by almost 0.3 dex. 
\label{w_vs_feh}}
\end{figure}

Further, Rutledge \etal (1997) have shown that the
non-linear relation between W$^{\prime}$ and [Fe/H]$_{\rm ZW84}$ becomes
linear when [Fe/H]$_{\rm CG97}$ is used instead.  This is illustrated in
Fig.\ \ref{w_vs_feh} where we plot the W$^{\prime}$ values for the
calibration clusters in Table 4 of DA95 against [Fe/H]$_{\rm CG97}$.  A
single straight line fits the data over the entire [Fe/H]$_{\rm CG97}$
abundance range (cf.\ Rutledge \etal 1997).  A least squares fit to these
data then yields:
\begin{equation}
{\rm [Fe/H]}_{\rm CG97} = (0.416 \pm 0.018) {\rm W}^{\prime} - (2.748 \pm 0.064)
\end{equation}
The rms deviation in [Fe/H]$_{\rm CG97}$ about the fitted line is 0.07 dex.
Rutledge \etal (1997) find a very similar relation.  Fig.\ \ref{w_vs_feh}
also shows the two segment calibration relation of DA95 for comparison.
At the moment it is not obvious which abundance scale is correct 
though the high dispersion studies of Sneden, Kraft and co-workers
(e.g., \markcite{KS95}Kraft \etal 1995 and references therein), for example,
do support the Carretta \& Gratton (1997) results for the clusters in common.

With the exception of Lindsay 11, all the observed SMC cluster W$^{\prime}$
values fall in the range where the (W$^{\prime}$, [Fe/H]$_{\rm ZW84}$) and
(W$^{\prime}$, [Fe/H]$_{\rm CG97}$) calibrations differ significantly.
Thus we will interpret our results using both calibration relations.
In Table \ref{abund_tab} then, we list 4 abundances for each SMC cluster.
These are abundances with and without the correction for the 
brightening of the horizontal
branch with decreasing age, and for both the [Fe/H]$_{\rm ZW84}$ calibration
of DA95 and the [Fe/H]$_{\rm CG97}$ derived above.  The errors listed 
include both the uncertainty in the W$^{\prime}$ value from 
Table \ref{results_tab} and the uncertainty in the particular calibration.
These uncertainties are comparable in size for clusters such as Lindsay~113,
but for clusters with smaller errors in W$^{\prime}$, such as NGC~339, the
calibration error dominates.
In general, the results in Table \ref{abund_tab} reveal no major surprises:
the abundances are not very different (though of higher precision)
from existing estimates based
on integrated spectroscopy and photometry and/or red giant branch colors.
For example, Zinn \& West (1984) list [Fe/H] = --1.51 $\pm$ 0.15 for
NGC~121 in excellent agreement with our value of 
[Fe/H]$_{\rm ZW84}$ = --1.46 $\pm$ 0.10 dex.

\section{Results} \label{results}

\subsection{Kinematics of the Cluster Sample} \label{motions}

The analysis of the kinematics of the cluster system of the LMC has
revealed generally ordered motion.  For example, the majority of the
clusters form a disk system whose parameters agree with those of the
optical isophotes and the \ion{H}{1} rotation curve 
(\markcite{BS92}Schommer \etal 1992).  Further, unlike the at most slowly
rotating, pressure supported, galactic halo globular cluster system, even
the oldest LMC star clusters have the kinematics of a disk system: the
rotation amplitude is comparable to that of the younger clusters and the
velocity dispersion is relatively small (Schommer \etal 1992).  For the
SMC, however, the situation is apparently quite different.  Kinematic
studies of both planetary nebula (\markcite{MD85}Dopita \etal 1985) and
carbon star (\markcite{HS89}Hardy \etal 1989, 
\markcite{DH97}Hatzidimitriou \etal 1997) samples have not revealed any
indications of systematic rotation, while the recent high resolution
study of the \ion{H}{1} in the SMC reveals a complex structure dominated by the
effects of expanding \ion{H}{1} shells, rather than a rotating disk-like
structure (\markcite{LS97}Stavely-Smith \etal 1997).  The on-going interaction
with the LMC (and with the Galaxy) undoubtedly also influences the
kinematics of SMC populations, particularly in the outer regions to north-east
of the SMC center and in the Wing
(e.g., \markcite{DH93}Hatzidimitriou \etal 1993).

While our sample of intermediate-age and old SMC star clusters is small,
it is nevertheless distributed over the entire spatial extent of
the SMC\@.  Thus it is worthwhile investigating the kinematics implied
by the cluster radial velocities.  Moreover, the ages of the star clusters 
studied range from
$\sim$4 to $\sim$12 Gyr and thus it is sensible to compare the cluster
results with those from the planetary nebula (Dopita \etal 1985) and carbon 
star (Hardy \etal 1989, Hatzidimitriou \etal 1997) samples.  Both these
populations have progenitors whose ages are comparable to those of the
star clusters.  Also available for comparison is the study of the
kinematics of SMC red clump stars
(Hatzidimitriou \etal 1993) and that for SMC field red giants 
(Suntzeff \etal 1986).  These studies are restricted to particular
areas; a region near NGC~121 in the latter case and a region approximately
3.5$\arcdeg$ north-east of the SMC center in the former case.

As mentioned above, both the planetary nebula and the carbon star 
samples show no evidence for any systematic rotation of the SMC, 
and our cluster velocities support this conclusion.  In particular,
the seven clusters fall within the scatter shown by the carbon stars in the
(V$_{\rm GSR}$, position angle) diagram of Hardy \etal (1989).  Similarly,
in the (V$_{\rm GSR}$, projected distance from the centroid along the 
SMC major axis) diagram of Dopita \etal (1985), the clusters are
indistinguishable from the planetary nebulae.  We have then used a maximum
likelihood code kindly provided by Tad Pyror to calculate the mean 
velocity $<{\rm V}_{\rm r}>$ and
velocity dispersion $\sigma$ for the cluster sample\footnote{For this sample,
the maximum likelihood analysis gives essentially identical results to those
generated by applying the formalism of 
\markcite{AD86}Armandroff \& Da~Costa (1986).}.
The results are, using heliocentric values, $<{\rm V}_{\rm r}>$ = 
138~$\pm$~6 km~s$^{-1}$ and $\sigma$ = 16 $\pm$ 4 km~s$^{-1}$.  The results
are summarized and compared with other samples in Table \ref{rv_data}.  The mean
velocity of the cluster sample is in good accord with those for the other 
samples.  The dispersion though, is somewhat lower: 
16 $\pm$ 4 km~s$^{-1}$ versus
21 -- 27 km~s$^{-1}$ for the other large area surveys.  Given the small
number of clusters in our sample we ascribe no significance to this difference.
It is interesting to note though that the 
dispersion of the old and intermediate-age
populations in the SMC is, at $\sim$24 km~s$^{-1}$, very comparable to
the dispersion seen for the LMC clusters of similar age (Schommer \etal 1992).
The major kinematic difference though lies in the substantial rotation of the
LMC cluster system (V$_{\rm circ}$ $\gtrsim$ 50 km~s$^{-1}$) which
is apparently
completely lacking in the SMC\@.  It is of further interest to note that 
Stavely-Smith \etal (1997) also find a similar dispersion (25 $\pm$ 0.6 km~s$^{-1}$) from their SMC \ion{H}{1} observations.  When combined with, 
for example, the
Cepheid observations of \markcite{MF88}Mathewson \etal (1988) for which
$\sigma$ = 22 $\pm$ 3 km~s$^{-1}$, it is evident that the kinematics of the
various SMC populations studied are all similar. As a result, it
appears that kinematics are independent of age in the SMC\@.  This 
result contrasts 
rather strongly with the situation in our Galaxy and in the LMC, but whether it
is a result of the interaction history of the SMC remains unclear.

\subsection{The SMC Age-Metallicity Relation} \label{amr}

In Table \ref{results_tab} we list our adopted ages for the six clusters
in our sample for which we have determined metallicities.  These ages 
have generally 
been taken from the original c-m diagram references (cf.\ Sect.\ \ref{samp}) 
and are appropriate for our assumed SMC distance
modulus of (m--M)$_{0}$ = 18.8.  For Kron~3, however, we have taken account of
the results of Alcaino \etal (1996) who found a somewhat older age than
did Rich \etal (1984).  Similarly, for Lindsay~113, we follow
\markcite{SD86}Seidel \etal (1987a) who noted that the apparent magnitude of
the red clump in this cluster, which lies at the extreme eastern edge of
the SMC, suggests that it is $\sim$0.2 -- 0.3 mag closer than the standard SMC
modulus.  Consequently, we have adopted the Seidel \etal (1987a) value of
6 $\pm$ 1 Gyr for the age of this cluster rather than the somewhat younger
value given by Mould \etal (1984).  In all cases we have assigned relatively
generous error bars to the adopted ages.
Fig.\ \ref{amr_plt} plots our spectroscopic abundance 
determinations for these six clusters against the adopted ages.  The upper
panel shows the abundances on the Zinn \& West (1984) scale while the lower
panel employs the Carretta \& Gratton (1997) abundance scale.  In both panels
the abundance plotted is that corrected for the variation with age of the
horizontal branch luminosity (cf.\ Table \ref{abund_tab}).  

In Fig.\ \ref{amr_plt} we
have also plotted additional data relevant to defining the age-metallicity
relation of the SMC\@.  We now briefly discuss these data.  First, in
Fig.\ \ref{amr_plt} points are plotted for five additional clusters: NGC~152,
NGC~330, NGC~411, NGC~419 and NGC~458.  For NGC~152 the abundance ([Fe/H] =
--0.8 $\pm$ 0.3) is from \markcite{PH81}Hodge (1981a) and is based 
on the giant branch
color.  The adopted age of 1.9 $\pm$ 0.5 Gyr comes from considering the 
magnitude difference between the main sequence turnoff and the red clump in
the c-m diagram of \markcite{PW81}Hodge (1981b); see 
\markcite{MD88}Mould \& Da~Costa (1988) for details.  For NGC~411 we adopted 
the age (1.8 $\pm$ 0.3 Gyr) given by \markcite{DM86}Da~Costa \& Mould (1986) 
for our adopted SMC modulus.  The abundance for this cluster ([Fe/H] = --0.84
$\pm$ 0.3 dex) also comes from the giant branch color: Da~Costa \& Mould (1986)
determine an abundance of --0.6 $\pm$ 0.3 dex relative to the galactic
cluster NGC~7789 for which \markcite{EF95}Friel (1995) lists [Fe/H] = --0.24
dex.  The age (1.2 $\pm$ 0.5 Gyr) and abundance ([Fe/H] = --0.7 $\pm$ 0.3)
adopted for NGC~419 follow from the c-m diagram isochrone fits of
\markcite{DR84}Durand \etal (1984) who also indicate that integrated
Washington system photometry of this cluster requires [Fe/H] $\gtrsim$ --1.0
dex.  These results are consistent with those of \markcite{DR83}Rabin (1983)
who showed from integrated spectra that NGC~419 and NGC~411 occupy essentially
identical locations in a (hydrogen line strength, calcium K line strength)
diagram.  For NGC~458, \markcite{PS88}Papenhausen \& Schommer (1988) list 
an age of 0.3 Gyr based on fitting a log(Z/Z$_{\sun}$) = --0.23 isochrone
from \markcite{DV85}VandenBerg (1985).  Given the uncertainty of abundance
determination via isochrone fits, we have in this case adopted a larger
error for the abundance.  In particular, the adopted error covers the
comment of \markcite{SC92}Stothers \& Chin (1992) that NGC~458 is likely
to have an abundance similar to that of the average SMC young population.  Our
adopted age error of $\pm$0.2 Gyr also encompasses the Stothers \& Chin (1992)
age estimate of $\sim$0.1 Gyr for this cluster.  Finally, we adopt
an abundance of [Fe/H] = --0.82 $\pm$ 0.10 for the young 
(age = 0.025 $\pm$ 0.015 Gyr, \markcite{CH95}Choisi \etal 1995) 
cluster NGC~330.  This abundance is
from the high dispersion spectroscopic study of 
\markcite{HS97}Hill \& Spite (1997, see also \markcite{MB95}Meliani \etal 1995 and the references therein).

Figure \ref{amr_plt} also includes relevant data for SMC field objects.  In
particular, we have plotted, at an adopted age of 13 $\pm$ 1 Gyr, the 
mean abundance found by 
\markcite{BD82}Butler \etal (1982) for 7 SMC RR~Lyrae stars in a field near
NGC~121.  Further, Suntzeff \etal (1986)
have studied both spectroscopically and photometrically a proper motion
selected sample of $\sim$30 SMC red giants, again in a field near NGC~121.
The mean abundance of these stars is [Fe/H] = --1.56 $\pm$ 0.06 and a real
abundance spread is apparently present in these data.  Fig.\ 9 of 
Suntzeff \etal (1986) suggests a total abundance range from perhaps
[Fe/H] $\approx$ --2.1 to [Fe/H] $\approx$ --1.2.  Here it is relevant to note
that the Suntzeff \etal (1986) results are on the Zinn \& West (1984) scale
and that both the upper limit on the abundance distribution and the mean
abundance would be raised if the Carretta \& Gratton (1997) scale was used
instead.  However, of all the additional results presented in 
Fig.\ \ref{amr_plt},
these are the only ones affected by the abundance scale difference.
As for the likely ages of the stars in the Suntzeff \etal (1986) sample,
individual determinations
are, of course, not possible but we can turn to the field region results
of Stryker \etal (1985).
These authors find that the field population near NGC~121 is dominated
by old stars (unlike the LMC) and they suggest an age range from perhaps 
8 to 14 Gyr is present.  The actual age distribution, however, is unknown.

Finally, we also show in Fig.\ \ref{amr_plt} representative values of the
present-day abundance in the SMC.  These determinations are from the
high dispersion spectroscopic studies of
\markcite{RB89}Russell \& Bessell (1989), [Fe/H] = --0.65 $\pm$ 0.06 from
8 F-type supergiants, \markcite{LL92}Luck \& Lambert (1992), 
[Fe/H] = --0.53 $\pm$ 0.05 from 7 Cepheids and supergiants, and 
\markcite{VH97}Hill (1997), [Fe/H] = --0.69 $\pm$ 0.05 from 6 K-type
supergiants.  These studies are consistent with each other and none provide
any compelling evidence for the existence of any substantial abundance
range among the youngest populations in the SMC\@.  \markcite{RD90}Russell
\& Dopita (1990) also find no significant metallicity spread among the
\ion{H}{2} regions of the SMC, in agreement with the earlier result of
\markcite{BP78}Pagel \etal (1978).

\section{Discussion}

\begin{figure}
\epsscale{0.59}
\plotone{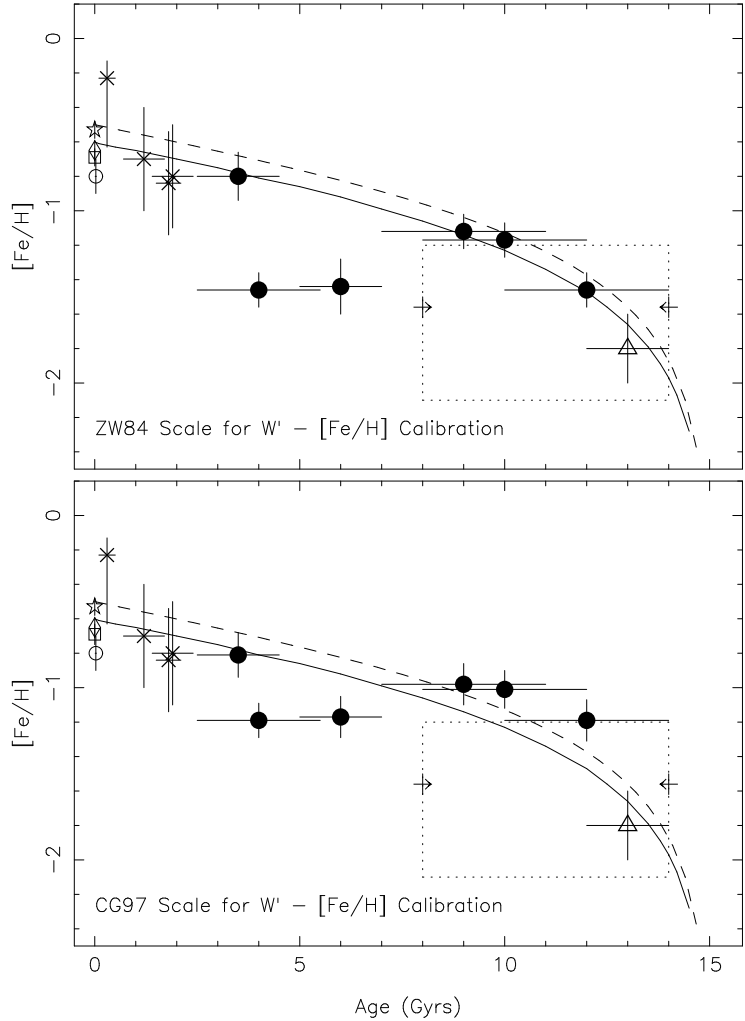}
\caption{The age-metallicity relation 
for the SMC\@.  Filled symbols
are the clusters observed in this paper with abundances on the Zinn \& West
(1984) scale shown in the upper panel and on the Carretta \& Gratton (1997)
scale in the lower panel.  The $\times$-symbols are the clusters, in order
of increasing age, NGC~458, NGC~419, NGC~411 and NGC~152, respectively, while
the open triangle is the SMC field RR Lyrae mean abundance of 
Butler \etal (1982).  The arrow symbols indicate the mean abundance of SMC 
field red giants from Suntzeff \etal (1986) and the box outlined by the
dotted lines indicates the range in abundance and age for these
stars.  The distribution of these stars within the box, however, is unknown.
The star-, diamond- and square-symbols are spectroscopic results for 
the present-day abundance of the SMC from Luck \& Lambert (1992), 
Russell \& Bessell (1989) and Hill (1997), respectively, while the open circle
is the cluster NGC~330 plotted using the abundance of Hill \& Spite (1997).
In both panels the solid and dashed lines are the relations predicted by 
simple chemical
evolution models.  Both curves use the same present-day gas mass fraction
(0.36) and formation epoch (15 Gyr) but the model depicted by the solid
line employs a present-day abundance of --0.6 dex, while that for the dashed
line uses --0.5 dex.
\label{amr_plt}}
\end{figure}

The appearance of the SMC age-metallicity relation in Fig.\ \ref{amr_plt}
is somewhat different from previous depictions of the cluster and field star
data (e.g., Stryker \etal 1985, \markcite{GD91}Da~Costa 1991, 
\markcite{OS96}Olszewski \etal 1996)
principally as a result of the increased precision of our spectroscopic
abundance determinations.  Leaving aside for the moment the two clusters
(Lindsay~113 and NGC~339) with anomalously low abundances, the enrichment
history for the SMC indicated by Fig.\ \ref{amr_plt} suggests a relatively
rapid ($\tau$ $\lesssim$ 3 Gyr) initial abundance increase followed by a
more modest rise starting at $\sim$10 Gyr and continuing until the present
day.  In particular, the previous requirement (e.g., Da~Costa 1991) for an
increased rate of enrichment to bring the abundance from [Fe/H] $\approx$ --1.3
at approximately 3 Gyr to the present-day value of [Fe/H] $\approx$ --0.6
is alleviated.  Indeed, the data of Fig.\ \ref{amr_plt}, again 
excepting Lindsay~113 and NGC~339, are now quite consistent with the 
predictions of the simple ``closed box'' model of chemical evolution.  
Thus it is also no longer necessary to postulate the significant gas infall or
strong galactic winds which were invoked (e.g., \markcite{MD91}Dopita 1991)
to explain the apparent lack of enrichment from $\sim$10 -- 12 Gyr to 
approximately 3 Gyr.  Of course such processes, however, may still take place.

The SMC, with its likely past interactions with the LMC and the Galaxy 
may well be an
``open'' rather than a closed box, but nevertheless simple models scaled
to a present-day abundance, the current SMC gas mass fraction (taken as
0.36, \markcite{JL84}Lequeux 1984) and a formation epoch, assumed to be at
15 Gyr, give reasonable representations of the data.  This is especially
the case in the upper panel of Fig.\ \ref{amr_plt} where we show simple models 
with present-day abundances of log~Z/Z$_{\sun}$ = --0.6 and
log~Z/Z$_{\sun}$ = --0.5 dex.  The
scatter about the model curves is entirely consistent with the uncertainties
in the data.  In the lower
panel of Fig.\ \ref{amr_plt} the same simple model curves are shown.  Here the
fit is somewhat less satisfactory: the rate of enrichment from $\sim$10~Gyr
to the present suggested by the observations being generally somewhat lower
than the model predictions.  Or conversely, the rate of enrichment prior to
$\sim$10 Gyr may have been more rapid than the simple model predicts.  The
model fits to the data in the lower panel of Fig.\ \ref{amr_plt} could no
doubt be improved by invoking infall, for example, but tighter constraints
on the age-metallicity relation are required before such additional 
calculations would be meaningful.  

We now turn to the ``anomalous'' clusters Lindsay~113 and NGC~339.   These
clusters
have abundances approximately 0.5 and 0.65 dex lower, respectively, than the simple model curves predict for their ages on the Zinn \& West (1984)
abundance scale, or $\sim$0.25 and 0.4 dex lower, respectively, using the 
Carretta \& Gratton (1997)
scale\footnote{In this latter case it could be argued that 
Lindsay~113 lies below
the model curve for a present-day abundance of --0.6 dex by an amount comparable to which clusters such as Lindsay~1 and Kron~3 lie above it.  Thus
Lindsay~113 would not be anomalous but would instead be a reflection of an
intrinsic abundance range of $\sim$0.2 dex present in the SMC 
over at least the 5 -- 10 Gyr epoch.  If it were not for NGC~339, which lies 
further below the model line
and which has an abundance difference of $\sim$0.4 dex on the Carretta \&
Gratton (1997) scale from the similar
aged cluster Lindsay~11, such an interpretation might be plausible.  However,
we prefer to work on the assumption that the bulk of the SMC population at 
any age less than $\sim$10 Gyr has only a small abundance range, and that
therefore these two clusters are anomalous.  Additional data will, of course,
eventually validate or refute this interpretation.}.
These clusters are considered anomalous since in dwarf
galaxies like the SMC, provided the star formation rate is relatively
constant and the infall of external material insignificant, the expanding
gas shells driven by evolving massive stars should throughly mix the
interstellar medium over galaxy-wide scales on timescales that are 
considerably less than a Hubble time (e.g., \markcite{RK95}Roy \& Kunth 1995,
see also \markcite{KS97}Kobulnicky \& Skillman 1997).
At the present epoch it does appear that the SMC is chemically homogeneous.
The studies listed above indicate that for both the young field stars and 
the \ion{H}{2} regions the maximum abundance dispersion permitted by the
observations is $\lesssim$0.1 dex.  Further, while there has been considerable
controversy in the past regarding the abundance of the young massive star
cluster NGC~330, recent work (e.g., Hill \& Spite 1997) suggests that this
cluster is only $\sim$0.1 dex more metal-poor than field objects of
comparable age, a minor offset.  As regards the LMC, where there is a large
population of intermediate-age star clusters, we can draw on the results of
\markcite{OS91}Olszewski \etal (1991) who also used \ion{Ca}{2} triplet
spectroscopy to determine abundances for a large number of LMC star clusters.
Olszewski \etal (1991) indicate that for the clusters in the age range 0.5 to
3 Gyr, the inner (r $<$ 5\arcdeg\ or $\sim$4 kpc) and outer clusters have
``apparently the same abundance spread, both of which are consistent with
the measurement errors (in) our metallicities''(Olszewski \etal 1991).  Thus
there is no compelling case for a significant abundance spread among the
LMC intermediate-age clusters.  However, in the Olszewski \etal (1991) sample,
one can find relatively well observed clusters such as Hodge~14 (5 observations
of 3 stars, [Fe/H] = --0.66) and NGC~1777 (3 observations of 3 stars, [Fe/H] =
--0.35) for which the results do suggest the existence of 
an intrinsic abundance range among the
intermediate-age LMC clusters of size perhaps $\lesssim$0.3 dex.  But this 
abundance range is considerably less than that between the SMC clusters
NGC~339 and Lindsay~11 ($\Delta$[Fe/H] = 0.66 $\pm$ 0.17 on the Zinn \& West
(1984) abundance scale, which is that used by Olszewski \etal 1991;
$\Delta$[Fe/H] = 0.38 $\pm$ 0.16 on the Carretta \& Gratton 1997 scale), again
pointing to the unusual nature of the two SMC low abundance clusters.  Indeed
these abundance differences are comparable in size to the {\it full range} in
abundance seen in the galactic disk age-metallicity relation at comparable
age (Edvardsson \etal 1993).

Can we then offer an explanation for their anomalously low abundances?  
We note first that
Lindsay~113 lies in the extreme eastern part of the SMC some 4.2\arcdeg\
($\sim$4.2 kpc in projection) from the center, while NGC~339 lies 1.5\arcdeg\
to the south.  This might suggest an abundance gradient but in fact all the
other clusters in our abundance sample are further from the SMC center, at 
least in projection, 
than NGC~339.  In particular, Lindsay~1 lies 3.4\arcdeg\ to the west and does
not appear anomalous.  We can also use the age-compensated horizontal branch
magnitudes to investigate any possible line-of-sight distance variations.
These data suggest that Kron~3, Lindsay~11, NGC~121 and NGC~339 are at
approximately the same distance, while Lindsay~1 and Lindsay~113 are perhaps
$\sim$0.3 mag closer.  Again it seems that a simple abundance gradient is
not the answer.

As for other explanations, a number of possibilities
exist.  Roy \& Kunth (1995), for example, suggest that large abundance
discontinuities might arise in dwarf galaxies lacking significant 
differential rotation if there are long dormant phases between successive
star forming episodes.  This might appear an attractive possibility given the
SMC's apparent lack of rotation, but the
existence of a relatively uniform age distribution for the SMC star
clusters with main sequence turnoff age determinations 
(cf.\ Fig.\ \ref{amr_plt}) argues against the required long dormant phases.

The remaining possibility is that the formation of these clusters involves
the infall of unenriched, or at least less enriched, gas\footnote{We postulate
that the gas from which these clusters formed consists of a mixture of SMC gas
with the abundance ``expected'' for the cluster ages, diluted by a component
with a lower abundance.  Given that these clusters have abundances Z 
approximately a factor of 3 lower than expected from the simple model curves,
the required dilution factors range from 0.67 for unenriched gas to 1.00
for gas with abundance equal to that of the clusters.}.  Given the 
extensive \ion{H}{1} halo surrounding both Magellanic Clouds 
(e.g., \markcite{MF84}Mathewson \& Ford, 1984) and the existence of the
Magellanic Stream, this possibility seems plausible.  We note, however,
that \markcite{LS94}Lu \etal (1994) have used absorption line spectroscopy
of a background source to argue that the Magellanic Stream gas is certainly
not of primordial (log~Z/Z$_{\sun}$ $\lesssim$ --2) composition.  Indeed
the abundance limits are consistent with present-day Magellanic Cloud
abundances (Lu \etal 1994).  If these abundances are generally applicable to
the Magellanic Stream gas, then it would seem to rule out the Magellanic
Stream as the source of the low abundance gas involved in the formation of
these clusters.  However, Lu \etal (1994) caution that the greatest
uncertainty in abundance studies of this type is the sampling differences
that can result from the narrow pencil beam of the absorption line data
versus the large beam of the 21cm data.  Since the \ion{H}{1} column density
can vary substantially on small scales (cf.\ the high resolution SMC
\ion{H}{1} map of Stavely-Smith \etal 1997 versus the earlier Parkes data
of, for example, Mathewson \& Ford 1984), the true abundances 
could differ substantially from the derived values.  We note also that 
postulating a Magellanic Stream origin for the low abundance gas involved
in the formation of these clusters requires that the Stream have an age of
at least 4 Gyr.  This conflicts with at least some models for the Stream's
origin (e.g., \markcite{GN96}Gardiner \& Noguchi 1996).

A further speculative possibility is that the
gas from which Lindsay~113 and NGC~339 formed (presumably separately since
the clusters have different ages) contained a low abundance component which
had been expelled in galactic wind at an early epoch, but which remained 
bound to the SMC and which subsequently cooled and fell back 
(cf.\ \markcite{TT96}Tenorio-Tagle 1996,
\markcite{BR97}Burkert \& Ruiz-Lapuente 1997).  A second speculative 
possibility is that the SMC acquired these clusters in a past interaction
much in the same way that the Galaxy is now acquiring clusters from the
Sagittarius dwarf galaxy (e.g., DA95).  However, we note that among the
dwarf galaxies of the Local Group, the only system known to contain 
star clusters of age similar to Lindsay~113 and NGC~339 (i.e., $\sim$4 -- 6 Gyr)
is the SMC itself!

We conclude by noting that the chemical evolution of the SMC, as exhibited
by the data described here for old and intermediate-age clusters, is
obviously quite complex.  If we are to provide further constraints on the
processes involved, then it is necessary to determine abundances (and ages)
for additional SMC star clusters.  There are a number of candidate clusters 
(e.g., NGC~361) whose ages are likely to lie between 1 to 2 Gyr and 
$\sim$10 Gyr.  The results for such clusters would help enormously in 
clarifying whether
the low abundances exhibited by Lindsay~113 and NGC~339 are common, or are
restricted to only a small fraction of SMC cluster population.  There 
are no technical difficulties preventing the collection of such data.

\acknowledgements

The authors wish to thank Dr.\ Taft Armandroff for reducing some of the 
spectra used in this paper and Dr.\ Tad Pyror for a copy of his maximum
likelihood velocity dispersion code.  We are also grateful to Dr.\ Ed 
Olszewski for supplying details of his study of NGC~339 in advance of
publication and to an anonymous referee for helpful comments.  The AAT staff
also deserve our thanks for support at the telescope.

\clearpage

\begin{deluxetable}{lrllcc}
\tablecolumns{6}
\tablecaption{Data for Program Cluster Stars \label{data_tab}}
\tablewidth{430pt}
\tablehead{
\colhead{Cluster}&\colhead{Star}&\colhead{$V$}&\colhead{$B-V$}& 
\colhead{W$_{8542}$+W$_{8862}$(\AA)}&\colhead{Notes}
}
\startdata
Lindsay 1 & OSA783 & 17.01 & 1.37 & $5.76 \pm 0.28 $ & 1\nl
          & OAS426 & 17.02 & 1.39 & $5.38 \pm 0.22 $ & 1\nl
          & OAS453 & 17.06 & 1.36 & $5.41 \pm 0.25 $ & 1\nl
          & OAS772 & 17.11 & 1.38 & $5.67 \pm 0.32 $ & 1\nl
\tablevspace{4pt}
Kron 3    & RDM406 & 16.69 & 1.51 & $6.23 \pm 0.42 $ & 2\nl
          & RDM408 & 16.80 & 1.45 & $5.86 \pm 0.41 $ & 2\nl
          & RDM303 & 16.83 & 1.44 & $6.01 \pm 0.33 $ & 2\nl
          & RDM099 & 17.25 & 1.28 & $5.39 \pm 0.43 $ & 2\nl
\tablevspace{4pt}
Lindsay 11 & MJD182 & 16.86 & 1.56 & $6.38 \pm 0.33 $ & 2\nl
           & MJD240 & 16.95 & 1.47 & $5.03 \pm 0.23 $ & 1,2,3\nl
           & MJD177 & 17.19 & 1.33 & $6.42 \pm 0.40 $ & 2\nl
           & MJD192 & 17.52 & 1.28 & $5.78 \pm 0.29 $ & 1,2\nl
           & MJD233 & 18.12 & 0.98 & $5.67 \pm 0.53 $ & 2,4\nl
           & MJD212 & 18.15 & 1.16 & $6.33 \pm 0.52 $ & 2,4\nl
\tablevspace{4pt}
NGC 121 & SDM029 & 17.03 & 1.40 & $4.71 \pm 0.33 $ & 2,5\nl
        & SDM063 & 17.34 & 1.34 & $4.74 \pm 0.44 $ & 2,4\nl
        & SDM123 & 18.04 & 1.09 & $4.43 \pm 0.44 $ & 2,4\nl
        & SDM132 & 18.06 & 1.14 & $4.55 \pm 0.32 $ & 2,4\nl
        & SDM193 & 18.12 & 1.16 & $5.01 \pm 0.27 $ & 2,4\nl
\tablevspace{4pt}
NGC 339 & OAS171 & 17.14 & 1.35 & $5.30 \pm 0.22 $ & 1\nl
        & OAS392 & 17.14 & 1.39 & $5.52 \pm 0.31 $ & 1\nl
        & OAS126 & 17.16 & 1.33 & $5.32 \pm 0.28 $ & 1\nl
        & OAS226 & 17.20 & 1.18 & $5.15 \pm 0.23 $ & 1\nl
        & OAS429 & 17.29 & 1.24 & $5.17 \pm 0.30 $ & 1\nl
        & OAS104 & 17.30 & 1.23 & $4.99 \pm 0.27 $ & 1\nl
\tablevspace{4pt}
NGC 361 & Arp I-15 & 16.4 & 1.4 & $6.08 \pm 0.32 $ & \nodata \nl
        & Arp I-35 & 16.5 & 1.4 & $5.82 \pm 0.33 $ & \nodata \nl
        & Arp I-51 & 16.7 & 1.4 & $6.20 \pm 0.28 $ & 1\nl
        & Arp I-63 & 16.8 & 1.2 & $5.45 \pm 0.30 $ & 1\nl
\tablebreak
\tablevspace{4pt}
Lindsay 113 & MDC280 & 16.96 & 1.32 & $5.20 \pm 0.36 $ & 2\nl
            & MDC098 & 17.03 & 1.28 & $6.28 \pm 0.62 $ & 2\nl
            & MDC281 & 17.04 & 1.17 & $5.28 \pm 0.26 $ & 2,6\nl
            & MDC078 & 17.21 & 1.23 & $4.42 \pm 0.44 $ & 2,4\nl
\tablevspace{4pt}
\enddata
\tablecomments{(1) Listed W$_{8542}$+W$_{8862}$ value is the 
mean from two spectra. (2) $V$, $B-V$ values derived 
from $R$, $B-R$ photometry.  (3) Line strength discrepant:
not a cluster member but radial velocity 
indicates SMC membership.  (4) Listed W$_{8542}$+W$_{8862}$ value 
is from
summed spectra.  (5) Radial velocity indicates Galactic foreground star.
(6) Listed W$_{8542}$+W$_{8862}$ value is the mean from three spectra.\\ 
{\it Identification and Photometry Sources}: 
Lindsay~1, Olszewski \etal (1987); Kron~3, Rich \etal (1984); 
Lindsay~11, Mould \etal (1992); NGC~121, Stryker \etal (1985);
NGC~339, Olszewski (priv.\ comm.); NGC~361, Arp (1958); Lindsay~113,
Mould \etal (1984).  Except for the photographic data of Arp (1958),
all photometry is derived from CCD images.}
\end{deluxetable}

\clearpage

\begin{deluxetable}{lccccc}
\tablecolumns{6}
\tablecaption{SMC Cluster Observed Data \label{results_tab}}
\tablewidth{430pt}
\tablehead{
\colhead{Cluster}&\colhead{Age}&\colhead{$V_{HB}$}&\colhead{W$^{\prime}$}
&\colhead{V$_{\rm r}$}&\colhead{N}\nl
&\colhead{(Gyr)}&&\colhead{(\AA)}&\colhead{(km s$^{-1}$)}
}
\startdata
Lindsay 1 & 10\phd\phn\phn$\pm$\phn2\phd\phn & 19.20  & 4.22 $\pm$ 0.11 & 126 $\pm$ 3 & 4  \nl
Kron 3 & \phn9\phd\phn\phn$\pm$\phn2\phd\phn & 19.44 & 4.29 $\pm$ 0.14 & 123 $\pm$ 3 & 4  \nl
Lindsay 11 & \phn3.5\phn$\pm$\phn1\phd\phn & 19.50 & 4.81 $\pm$ 0.18 & 132 $\pm$ 5 & 5  \nl
NGC 121 & 12\phd\phn\phn$\pm$\phn2\phd\phn & 19.60 & 3.74 $\pm$ 0.18 & 138 $\pm$ 4 & 4 \nl
NGC 339 & \phn4\phd\phn\phn$\pm$\phn1.5 & 19.36 & 3.91 $\pm$ 0.07 & 117 $\pm$ 8 & 6 \nl
NGC 361 & \nodata & \nodata & \nodata  & 164 $\pm$ 6 & 4 \nl
Lindsay 113 & \phn6\phd\phn\phn$\pm$\phn1\phd\phn & 19.13 & 3.91 $\pm$ 0.19 & 162 $\pm$ 4 & 4  \nl
\enddata
\end{deluxetable}

\clearpage

\begin{deluxetable}{lcccc}
\tablecolumns{5}
\tablecaption{SMC Cluster Abundance Results \label{abund_tab}}
\tablewidth{430pt}
\tablehead{\colhead{Cluster}&\multicolumn{2}{c}{[Fe/H]$_{\rm ZW84}$}&\multicolumn{2}{c}{[Fe/H]$_{\rm CG97}$}\nl
&\colhead{Raw}&\colhead{$V_{HB}$(age)$_{corrected}$}&\colhead{Raw}
&\colhead{$V_{HB}$(age)$_{corrected}$}
}
\startdata
Lindsay 1   & -1.14 $\pm$ 0.10 & -1.17 $\pm$ 0.10 & -0.99 $\pm$ 0.11 & -1.01 $\pm$ 0.11 \nl
Kron 3      & -1.08 $\pm$ 0.12 & -1.12 $\pm$ 0.12 & -0.96 $\pm$ 0.12 & -0.98 $\pm$ 0.12 \nl
Lindsay 11  & -0.70 $\pm$ 0.14 & -0.80 $\pm$ 0.14 & -0.75 $\pm$ 0.13 & -0.81 $\pm$ 0.13 \nl
NGC 121     & -1.46 $\pm$ 0.10 & -1.46 $\pm$ 0.10 & -1.19 $\pm$ 0.12 & -1.19 $\pm$ 0.12 \nl
NGC 339     & -1.36 $\pm$ 0.10 & -1.46 $\pm$ 0.10 & -1.12 $\pm$ 0.10 & -1.19 $\pm$ 0.10 \nl
Lindsay 113 & -1.37 $\pm$ 0.16 & -1.44 $\pm$ 0.16 & -1.12 $\pm$ 0.12 & -1.17 $\pm$ 0.12 \nl
\enddata
\end{deluxetable}

\clearpage

\begin{deluxetable}{lrcccc}
\tablecolumns{6}
\tablecaption{Kinematics of SMC Old and Intermediate-Age Populations \label{rv_data}}
\tablewidth{430pt}
\tablehead{\colhead{Sample}&\colhead{N}&\colhead{$<{\rm V}_{\rm r}>_{helio}$}&
\colhead{$<{\rm V}_{\rm r}>_{LSR}$}&\colhead{Dispersion}&\colhead{Ref}\nl
&&\colhead{(km s$^{-1}$)}&\colhead{(km s$^{-1}$)}&\colhead{(km s$^{-1}$)}
}
\startdata
Star Clusters & 7 & 138 $\pm$ 6 & 127 $\pm$ 6 & 16 $\pm$ 4 & 1 \nl
Planetary Nebulae & 44 &\nodata & 134 $\pm$ 4 & 25 $\pm$ 3 & 2 \nl
Carbon Stars\tablenotemark{a} & 131 & 148 $\pm$ 2 & 140 $\pm$ 2 & 27 $\pm$ 2 & 
 3 \nl
Carbon Stars\tablenotemark{b} & 62 &146 $\pm$ 3 & \nodata & 21 $\pm$ 2 & 4 \nl
Red Clump Stars\tablenotemark{c} & 29 &151 $\pm$ 6 & \nodata & 
33 $\pm$ 4\tablenotemark{d} & 5 \nl
Red Giants\tablenotemark{e} & 12 & 123 $\pm$ 7 & \nodata 
& 18 $\pm$ 5\tablenotemark{f} & 6 \nl
\enddata

\tablenotetext{a}{Central Region of SMC only}
\tablenotetext{b}{Principally Outer Regions of SMC, Outer Wing excluded}
\tablenotetext{c}{40$\arcmin$ diameter region 3.5$\arcdeg$ north-east
of the SMC center}
\tablenotetext{d}{Dispersion is $\sim$18 km s$^{-1}$ after
correction for velocity, distance correlation}
\tablenotetext{e}{Near NGC 121 only}
\tablenotetext{f}{Value is that corrected for observed velocity errors}
\tablecomments{{\it References}: (1) This paper, (2) Dopita \etal (1985),
(3) Hardy \etal (1989), (4) Hatzidimitriou \etal (1997), (5) Hatzidimitriou \etal (1993), (6) Suntzeff \etal (1986).}
\end{deluxetable}

\clearpage

\clearpage

\clearpage

\clearpage

\end{document}